\documentclass[conference]{IEEEtran}
\usepackage[linesnumbered, ruled, vlined]{algorithm2e}
\usepackage{graphicx}
\usepackage{balance}
\usepackage{caption}
\usepackage{moresize}
\usepackage{pbox}
\usepackage{mathtools}
\usepackage[TABBOTCAP]{subfigure}
\usepackage{paralist}
\usepackage[T1]{fontenc}
\usepackage{balance}
\usepackage[dvipsnames,table,xcdraw]{xcolor}
\usepackage{multirow}
\usepackage{multicol}
\usepackage{amsmath}
\usepackage{listings}
\usepackage{setspace}
\usepackage{verbatim}
\usepackage{float}
\usepackage{xspace}
\usepackage{amssymb}
\usepackage{ifthen}
\usepackage{pifont}
\usepackage{textcomp}
\usepackage{pict2e,picture}
\usepackage{url}
\usepackage{cite}
\usepackage{epsfig}
\usepackage[subfigure]{tocloft}
\usepackage{amsfonts}
\usepackage{amssymb}
\usepackage{latexsym}
\usepackage{booktabs}
\usepackage{tabularx}
\usepackage{rotating}
\usepackage[numbers]{natbib}
\usepackage{mdwlist}
\usepackage{colortbl}
\usepackage[normalem]{ulem}
\usepackage{tikz}

\include{macro}
\begin{document}
\title{Automating Software Development for \\ Mobile Computing Platforms}

\author{\IEEEauthorblockN{Kevin Moran}
\IEEEauthorblockA{Department of Computer Science\\
College of William \& Mary\\
Williamsburg, Virginia\\
Email: kpmoran@cs.wm.edu}}

\maketitle

\begin{abstract}
Mobile devices such as smartphones and tablets have become ubiquitous in today's modern computing landscape. The applications that run on these mobile devices (often referred to as ``apps'') have become a primary means of computing for millions of users and, as such, have garnered immense developer interest. These apps allow for unique, personal software experiences through touch-based UIs and a complex assortment of sensors. However designing and implementing high quality mobile apps can be a difficult process. This is primarily due to challenges unique to mobile development including change-prone APIs and platform fragmentation, just to name a few.  This paper presents the motivation and an overview of a dissertation which presents new approaches for automating and improving mobile app design and development practices. Additionally, this paper discusses potential avenues for future research based upon the work conducted, as well as general lessons learned during the author's tenure as a doctoral student in the general areas of software engineering, maintenance, and evolution.
\end{abstract}

\IEEEpeerreviewmaketitle

\section{Dissertation Overview~~\cite{Moran:Dissertation'18}}
\label{ch1:sec:overview}

\begin{quote}
\begin{singlespace}
	\textit{``The essence of a software entity is a construct of interlocking concepts: data sets, relationships among data items, algorithms, and invocations of functions. This essence is abstract in that such a conceptual construct is the same under many different representations. It is nonetheless highly precise and richly detailed. I believe the hard part of building software to be the specification, design, and testing of this conceptual construct, not the labor of representing it and testing the fidelity of the representation.''}
\end{singlespace}

\begin{flushright}
    \textit{\footnotesize{-- Fredrick Brooks, No Silver Bullet -- Essence and Accident in Software Engineering (1986)}}
  \end{flushright}
\end{quote}

\begin{figure}
\centering
\vspace{-0.2cm}
\includegraphics[width=0.87\columnwidth]{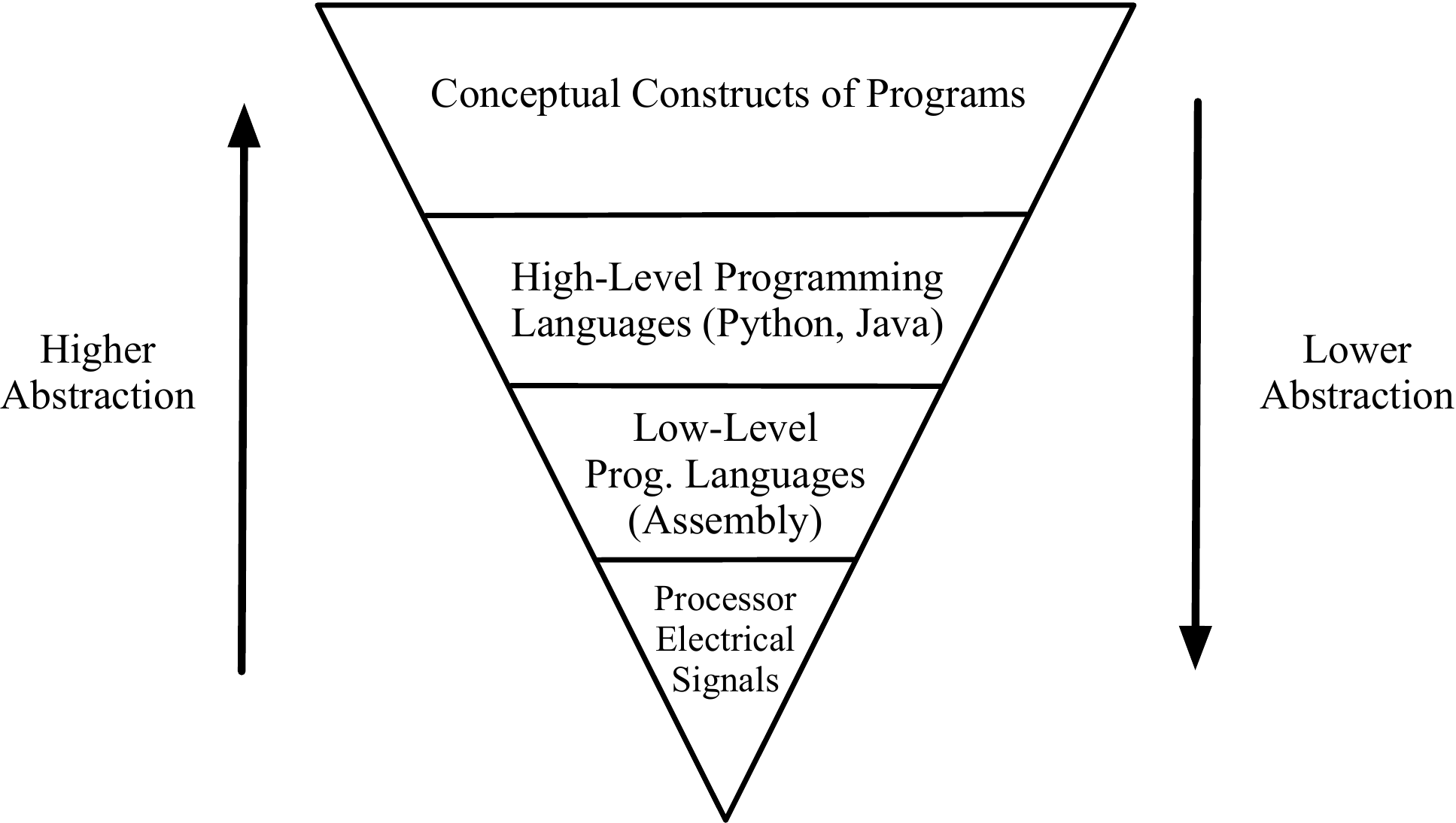}
\centering
\caption{The Hierarchy of Abstraction in Computer Science}
\label{fig:abstractions}
\end{figure}

	Software developers inherently reason about several different abstractions of ideas. In fact, the foundations of computer science more broadly are centered upon a hierarchy of abstractions (Fig. \ref{fig:abstractions}). This hierarchy begins at the lowest level with the physical representation of computers as a complex assortment of electrical signals, moves to representations of ideas in code that are able to carry out logical processes using these signals, and culminates at the highest level in mental models of logical processes for solving problems. At its core, computer science is largely concerned with the interplay between the various levels of this abstraction hierarchy. In his widely regarded \textit{``No Silver Bullet''} essay~\cite{Brooks:Computer'87} Fredrick Brooks identifies two key abstractions that modern software engineers face: (i) \textit{conceptual constructs} (\ie mental models of a given development task), and (ii) \textit{representations} of these conceptual constructs (\ie their concrete instantiation in a medium such as code, or natural language). Brooks argues that the most critical part of the software development process is not the transferral of concepts into a concrete representation, but rather the conceptualization of interlocking constructs that inherently constitute a piece of software.  Although translating abstract mental models into a tangible artifact like code is not a trivial process, Brooks recognized that the mental formulation of \textit{what} needs to be built is the most crucial step, as it is a distinctly abstract process requiring intellectual acuity. This is logically evident, as a faithful representation of ineffective ideas in code inevitably  results in an unsuccessful program. This conceptualization of the mental model of a program is what Brooks refers to as the \textit{essence} of software engineering. 

	With this view of the software development process in mind, there are evidently two major courses of research for improving engineering practices. Namely, conducting studies to understand, and designing techniques to aid and automate (i) the derivation of a conceptual model embodying the requirements, specifications, and design of a software system, and (ii) the process of translating this conceptual model into a concrete representations that can be understood, executed, and maintained by both humans and computers. However, directly addressing this first course of research is exceedingly difficult, as it attempts to directly operate upon the \textit{essence} of software engineering. As Brooks argues, due to the widely variable nature of software projects and the distinctly unique thought processes of humans, there is unlikely a ``silver bullet'' that dramatically improves the process of \textit{conceptualizing} software. 
 
	Humans are likely to play a major role in the process of software development for the foreseeable future, and as such the process of developing conceptual constructs of software is likely to persist.  Thus, in order to move the field forward, an appropriate avenue of research relates to the second course outlined above and involves helping to make the instantiation of conceptual software constructs as frictionless as possible. In fact, such work directly targets Brook's prescriptions for dealing with the difficulties that arise related to the \textit{essence} of software engineering namely, \textit{rapid prototyping and iteration} and \textit{growing software organically}~\cite{Brooks:Computer'87}.

	The work presented in this dissertation attempts to facilitate the process of instantiating conceptual software development concepts into accurate, effective representations through automation. The hope is that by automating different parts of the software design, development and testing processes, we will be able to allow developers to focus more effectively on the important task of conceptualizing the data, algorithms and functions that underlie the problem or task to which the software will be applied; thus facilitating the rapid iteration and organic evolution of intuitive, elegant programs.

\section{Motivation - Software Language Dichotomies}
\label{ch1:sec:Motivation}

	As explained in the previous section, the work conducted in the presented dissertation is aimed at designing techniques to \textit{automate} various parts of the software development process. This automation is meant to facilitate the instantiation of conceptual constructs of software into concrete representations.  However, these concepts can be concretely represented in several different manners, such as code, natural language, or in graphical user interfaces (GUIs). In this section, we further motivate the work conducted by examining development challenges that surface as a result of the interplay between different representations of software.

	Specific challenges in software engineering often stem from difficulties navigating different pairs of languages. For instance, when considering challenges related to software traceability, developers must reason between program representations related to natural language and code, interpreting how concepts and functional specifications dictated in natural language are dispersed throughout a codebase.  When designing the graphical user interface of program, designers and developers must reason between the modalities of code and pixel-based image representations of the app via the graphical user interface.  These pairs of contrasting information modalities have been labeled as \textit{language dichotomies}~\cite{Moran:ICPC'18}.  Developing solutions to help developers more effectively reason between various language dichotomies is a key factor in helping to overcome many program comprehensions challenges.

	More specifically, a language dichotomy can be defined as \textit{a difficulty in program comprehension resulting from reasoning about different representations or modalities of information that describe a program}~\cite{Moran:ICPC'18}.  There are several language dichotomies that contribute to a varied set of problems.  In the presented dissertation we focus on three different modalities of information:

\begin{enumerate}
	\item{\textbf{\textit{Natural Language}:} This modality represents languages that humans typically use to convey ideas or information to one another, such as English.}
	\item{\textbf{\textit{Code}:} This modality represents the languages that humans utilize to construct a program, such as Java or Swift.}
	\item{\textbf{\textit{Graphical User Interfaces (GUIs)}:} Much of today's user facing software is graphical, and mobile apps are no exception.  This information modality is highly visual, consisting of pixel-based representations of a program typically comprised of a logical set of building blocks often referred to as GUI-widgets or GUI-components.}
\end{enumerate}

	We assert that these language dichotomies can be effectively bridged through \textit{automation}, thus helping to overcome several resulting software development problems. More specifically, by building upon techniques related to program analysis, machine learning, and computer vision, techniques can be derived to help automatically translate information across modalities, or detect anomalies between corresponding program representations in a single modality. 

\section{Research Context: Mobile Applications}
\label{ch1:sec:Motivation}

	In order to devise new approaches that automate the various components of the software development lifecycle, we need a suitable domain within which we can instantiate and evaluate them.  In the scope of the presented dissertation, we focus our efforts on \textit{mobile applications}.  Mobile applications, often referred to colloquially as ``apps'', are quite simply software applications that run on mobile hardware such as smartphones or tablets.  Choosing mobile applications as our research domain is beneficial for at least the following three reasons: (i) there are open challenges unique to the software development process for mobile apps, providing a fertile research landscape, (ii) mobile apps, and by extension mobile app development, are \textit{extremely} popular, giving our work a large potential for practical impact, and (iii) mobile platforms provide a wide array of frameworks and utilities that facilitate varying types of program analysis. It should be noted that the work carried out in the presented dissertation is instantiated for the Android platform, mainly due to its open source nature and the litany of supporting tools and frameworks surrounding the platform. However, there are no substantial technical barriers that prevent the techniques presented in this dissertation from being transferred to other platforms.

\section{Summary of Contributions}
\label{ch1:sec:context}

	The core premise of presented dissertation is as follows:

\begin{quote}
\begin{singlespace}
	\textit{Automating the process of instantiating and reasoning about concrete representations of conceptual software constructs in code, natural language, and graphical user interfaces allows for more effective software development by enabling rapid prototyping, fast iteration, and by allowing software to grow organically as abstract concepts evolve.}
\end{singlespace}
\end{quote}

	To prove out this thesis, we develop models and approaches for automating the design, implementation, and testing of mobile applications. In particular, we develop novel approaches for automatically constructing and testing the Graphical User Interface (GUI) of mobile apps through novel applications program analysis, computer vision, and machine learning techniques. We illustrate that the techniques presented in this dissertation represent significant advancements in mobile development processes through a series of empirical investigations, user studies, and industrial case studies that demonstrate the effectiveness of these approaches and the benefit they provide developers.

\subsection{Automated Reporting of GUI Design Violations for \\Mobile Apps}

\begin{figure}
\centering
\vspace{-0.4cm}
\includegraphics[width=0.85\columnwidth]{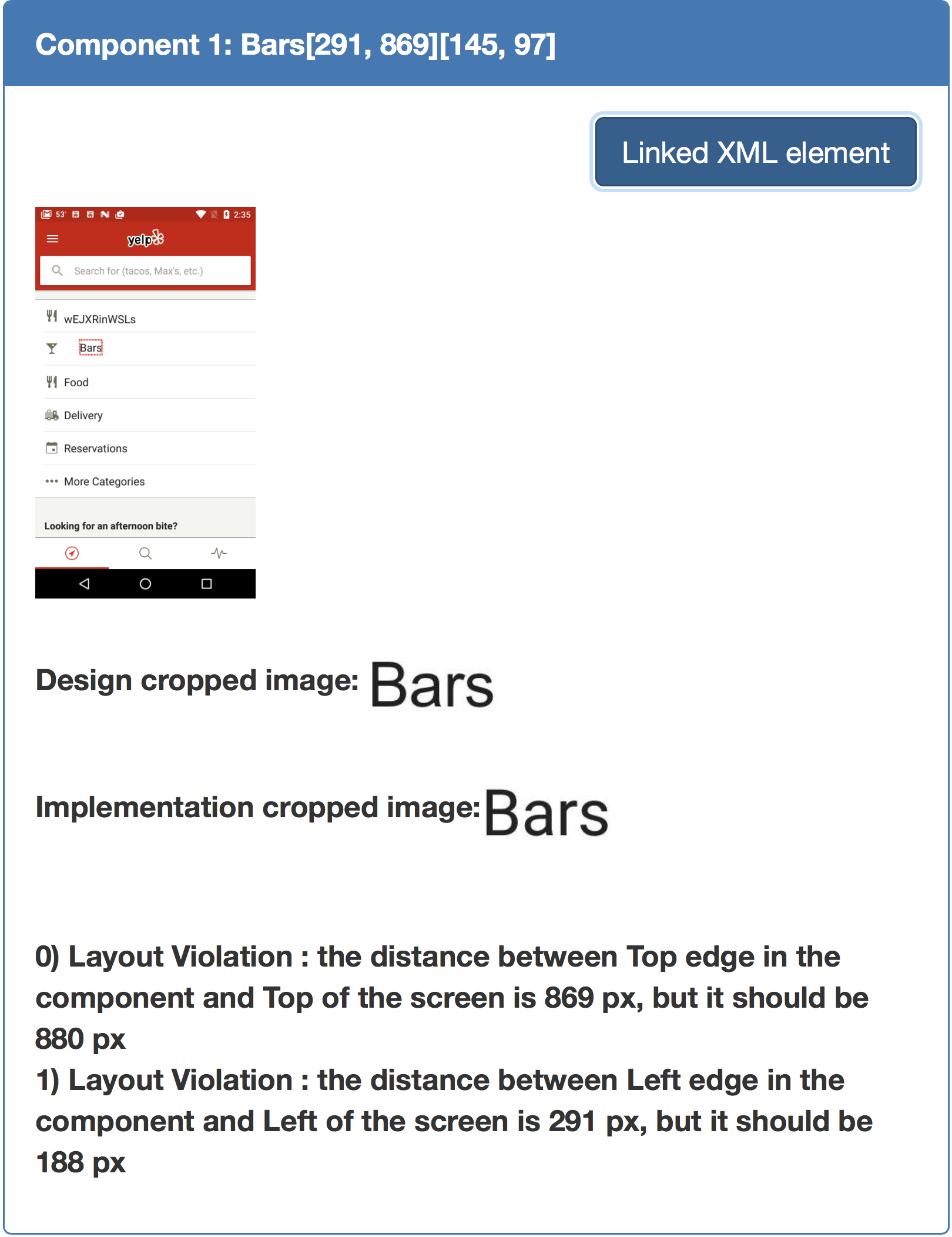}
\centering
\caption{Example GUI Design Violation from \GVT.}
\label{fig:gvt}
\end{figure}

	First, we develop a new technique, called \GVTsp, for detecting and reporting instances where the GUI of a mobile application does not adhere to its intended design specifications as stipulated in a mock-up. This technique receives as input two images with accompanying metadata, one for the mock-up and one screenshot of the implemented GUI, and generates a detailed report stipulating instances where the specifications of the mock-up were not properly implemented.  Our approach generates hierarchal models of the mock-up and implementation of a particular screen of an application's GUI and relates this model to the pixel-based images using coordinates.  It then applies a computer vision technique for measuring perpetual differences in images modeled after the human visual system, and categorizes image differences according to an empirically derived taxonomy of GUI implementation errors (See Figure \ref{fig:gvt} for example report). To evaluate \GVTsp we carried out both a controlled empirical evaluation with open-source applications as well as an industrial evaluation with designers and developers from Huawei, a major software and telecommunications company. The results show that \GVTsp is able to detect and report violations of GUI design specifications with remarkable efficiency and accuracy and is both useful and scalable from the point of view of industrial designers and developers. \GVTs industrial applicability is bolstered by the fact that, at the time of this dissertation's publication, over one-thousand industrial designers and developers at Huawei actively utilize our approach to improve the quality of their mobile apps. This work is based primarily on the paper from Moran, Li, Bernal-Cardenas, Jelf, and Poshyvanyk~\cite{Moran:ICSE'18}.  

\subsection{Machine Learning-Based Prototyping of Graphical User Interfaces for Mobile Apps}

\begin{figure}
\centering
\vspace{-0.4cm}
\includegraphics[width=\columnwidth]{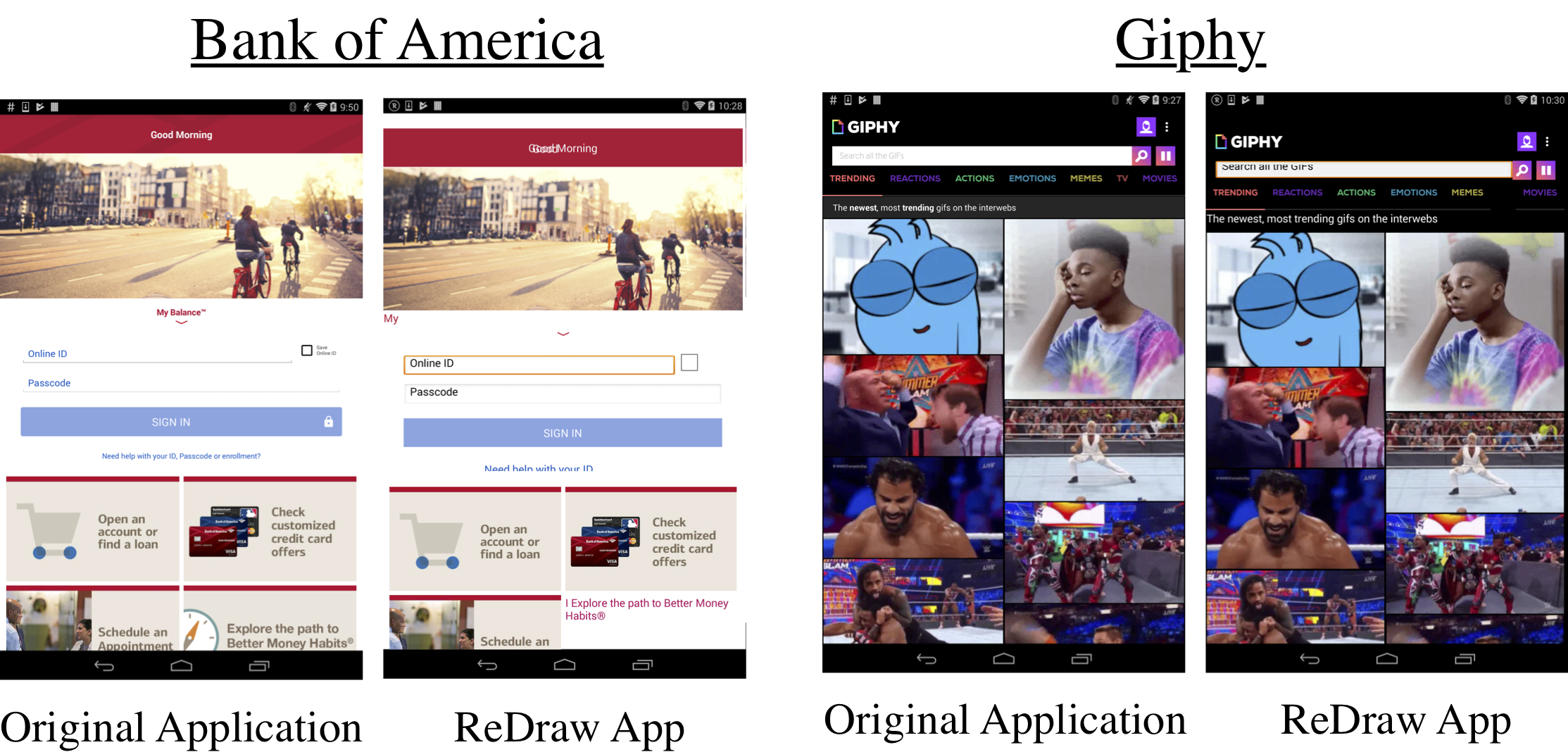}
\centering
\caption{Prototype Applications Generated by \ReDraw}
\label{fig:redraw}
\end{figure}

	Second, we devise a technique, called \ReDraw, to automate the process of translating an image-based mock-up of a mobile application's GUI into suitable code.  Our approach decomposes this translation process into three major steps: (i) detection of GUI elements, (ii) classification of these GUI elements into domain-specific, programmatic categories, and (iii) the construction and assembly of these categorized GUI elements into hierarchical code representation. Techniques from computer vision are utilized to \textit{detect} GUI elements in image-based representations of GUIs.  A deep convolutional neural network (CNN) trained on automatically derived image labels from tens of thousands applications screens is utilized to accurately \textit{classify} GUI elements into programmatic categories. Finally, we develop a data-driven k-nearest neighbors (KNN) algorithm for constructing realistic hierarchical representations of an app's GUI before translating this representation into code.  Our evaluation of \ReDraw illustrates that our approach's CNN achieves an average GUI-component classification accuracy of 91\% and assembles prototype applications that closely mirror target mock-ups in terms of visual affinity while exhibiting reasonable code structure (See Figure \ref{fig:redraw}). Furthermore, interviews with industrial practitioners from Google, Facebook, and Huawei illustrate \ReDraws potential to improve real design and development workflows. This technique is based primarily on work by Moran, Bernal-Cardenas, Curcio, Bonett, and Poshyvanyk~\cite{Moran:ArX'18}. 

\subsection{Improving GUI-based Testing for Mobile Applications}

\begin{figure}
\centering
\includegraphics[width=\columnwidth]{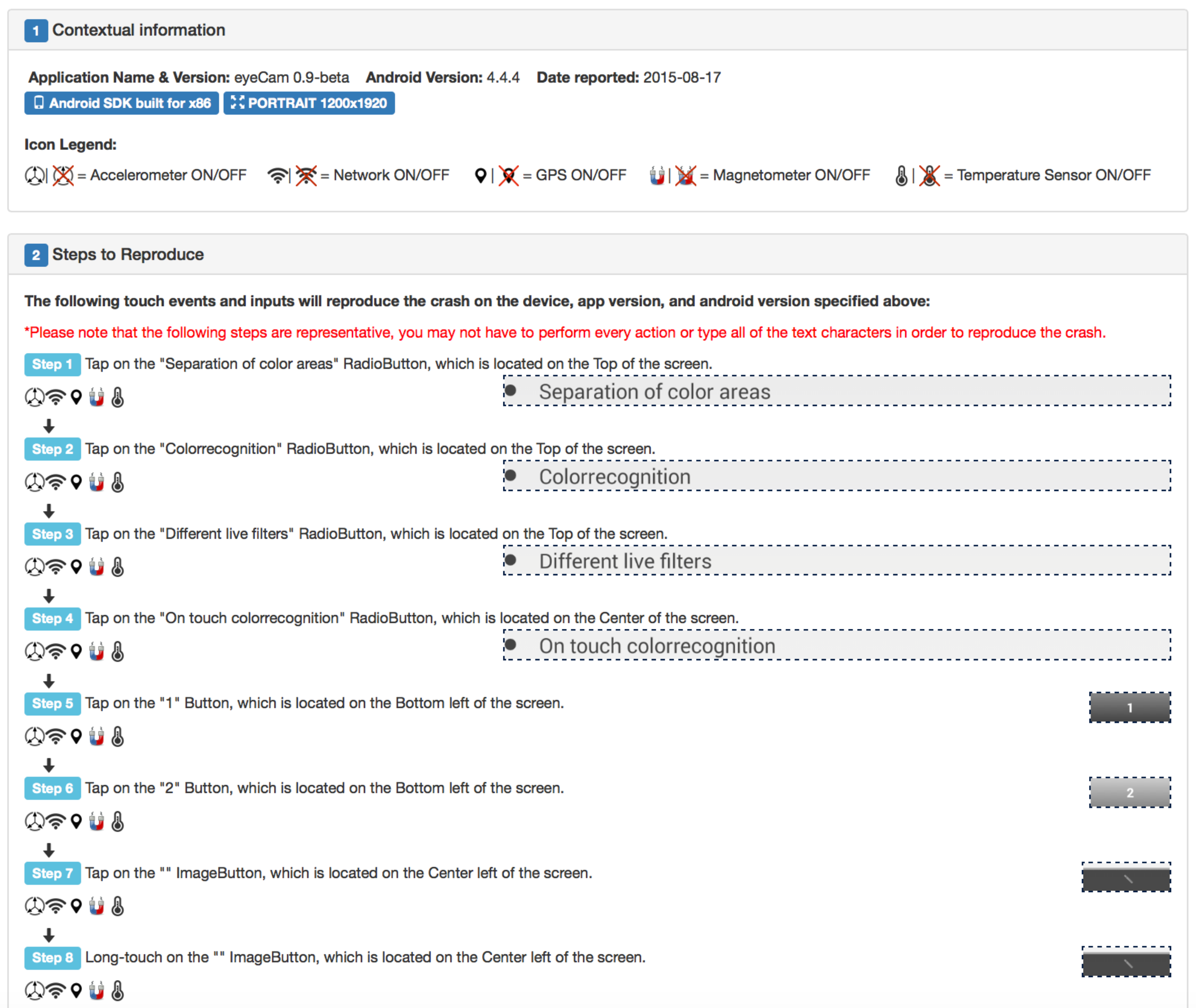}
\centering
\caption{Example Section of a \CrashScope Report}
\label{fig:crashscope}
\end{figure}

	Finally, we present a novel approach, called \CrashScope for automated testing of mobile applications. Our approach develops a new technique for constructing an on-the-fly event flow model of an application using systematic GUI exploration. Furthermore, our approach is capable of analyzing an application both statically and dynamically, extracting program features more likely to induce crashes, and stress-testing these features according to one of several strategies. We evaluated \CrashScopes effectiveness in discovering crashes as compared to five state-of-the-art Android input generation tools on 61 applications. The results demonstrate that \CrashScope is able to uncover crashes that other tools failed to detect and provides more detailed fault information. Additionally, in a study analyzing eight real-world Android app crashes, we found that \CrashScopes reports are easily readable (See Figure \ref{fig:crashscope}) and allow for reliable reproduction of crashes by presenting more explicit information than human written reports. This approach is based primarily on work related to the \CrashScope, {\small \sc Fusion}, and {\small \sc MonkeyLab} approaches conducted by Moran, Linares-Vasquez, Bernal-Cardenas, Vendome, White, and Poshyvanyk~\cite{Moran:ICST'16,Moran:ICSE-C'17, Moran:FSE'15, Moran:ICSE'16, Linares-Vasquez:MSR'15, Moran:MobileSoft'17}.

\section{Future Research Vision}

	The dissertation described in this paper presents several different approaches for automating the software development process of mobile applications. More specifically, we have helped to automate various aspects of the design, implementation, and testing of apps.  However, the presented work only touches the surface of various components of the development process that are ripe for automation.  Thus, there are several promising avenues of future work related to software development automation. In this chapter, we outline three such topics.

\subsection{Short Term: Toward Automatically Documenting GUIs}

	Modern software development practices in nearly all domains are tightly constrained by aggressive release deadlines and pressure to continuously evolve software over time. This situation is worsened by the fact that developers often spend a majority of their development time comprehending code~\cite{Ko:ICSE'05,Roehm:ICSE'12}. As outlined in Section \ref{ch1:sec:Motivation}, many of the challenges that developers face in comprehending code can be traced back to difficulties in reasoning between different abstractions of software, such as screenshots and functional code, or bug reports and code~\cite{Moran:FSE'15,Moran:ICPC'18}. Thus, one promising avenue for future work lies in automatically documenting different aspects related to the graphical user interface of GUI-based software applications. This work could lie in automatically summarizing changes as an app evolves over time, or in automatically documenting test cases based on information inferred from the GUI of screens that are tested.

The overarching goals of this future research thrust regarding automated GUI documentation are as follows:

\begin{itemize}

	\item{\textbf{Goal 1:}\textit{ Understanding Developer's and User's Information Needs in Documenting GUIs:} In order to create effective automated documentation for graphical user interfaces, it is important to first understand what documentation information both developers and users find useful. Thus, the first goal of this research thrust is to conduct studies that will shed light on information needs for GUI documentation.}

	\item{\textbf{Goal 2:}\textit{ Designing  Developer and User-Centric Approaches for Automated GUI Documentation:} Once we have established a set of guidelines for effective GUI documentation in eyes of developers and users, we will leverage this knowledge to create approaches that are capable of automatically documenting GUIs as software evolves.}
	
\end{itemize}

\subsection{Long-Term: Toward GUI-centric Automated Program Understanding \& Synthesis}

	The Graphical User Interfaces of software applications contain a wealth of information that may be useful for aiding in automated program understanding, and in the future, program synthesis. In this dissertation, we have illustrated an effective and promising approach for automated synthesis of code that implements a specified graphical user interface of a mobile app.  However, this is only the first step toward a more complete process of program synthesis. Given recent advancements in artificial intelligence and machine learning techniques, particularly as they relate to computer vision, one could conceive of moving beyond the capabilities of \ReDraw, towards implementing different functional properties of a GUI.  However, to accomplish this, the extent to which the visual semantics of graphical user interfaces can encode underlying functional information of a software GUI must be thoroughly explored. 

The overarching goals of this future research thrust in program understanding and synthesis are as follows:

\begin{itemize}

	\item{\textbf{Goal 1:}\textit{ Explore the Representational Power of Graphical User Interfaces:} In order to move toward approaches capable of automatically generating functional GUI-related code, the degree to which this functional information can be learned from GUIs must be explored. In essence, this requires studies focused on ascertaining the representational properties of GUIs as related to software functionality.}

	\item{\textbf{Goal 2:}\textit{ Designing Approaches for Synthesizing functional GUI-related Code:} According to the information gleaned from studying the representational power of GUIs, we will design approaches for synthesizing code related to various discrete functional properties of software GUIs.}
	
\end{itemize}

\subsection{Long Term: Toward a Practical, Comprehensive Framework for Automated GUI-based Testing}

GUI-based testing is a critical component of the of the testing pipelines of some of the most prevalent types of modern software, such as mobile apps.  There has been a wide range of research that aims to help automate various parts of the GUI-based testing process, with a particular focus on mobile apps~\cite{Linares-Vasquez:ICSME'17, Choudhary:ASE'15}.  However, many of these approaches have not yet been adapted in practice due to several practical shortcomings~\cite{Joorabchi:ESEM'12,Choudhary:ASE'15,Kochhar:ICST'15,Linares-Vasquez:ICSME'17a}.  Thus, it is important that research on GUI-based testing continues with a focus on tailoring automated approaches toward practical workflows commonly employed by modern developers. As future work, we aim to help bring about a paradigm shift in the thinking related to GUI-based testing, based upon the research vision  of CEL testing for mobile apps introduced by Linares-Vasquez, Moran, and Poshyvanyk~\cite{Linares-Vasquez:ICSME'17}. CEL mobile testing~\cite{Linares-Vasquez:ICSME'17} is founded on three principles aimed to overcome the shortcomings of current mobile GUI-based testing practices: \emph{Continuous}, \emph{Evolutionary}, and \emph{Large-scale} (CEL). At a high level, these principles dictate that (i) mobile apps should be tested \textit{continuously} according to different goals, (ii) testing artifacts should change in an \textit{evolutionary} fashion according to the often rapid development process of mobile apps, and (iii) GUI-based testing must be performed at a \textit{large-scale} to keep pace with software evolution and the ever growing landscape of mobile devices. 

To achieve the vision of CEL testing, we aim to focus on the following three research thrusts adapted from~\cite{Linares-Vasquez:ICSME'17}:

\begin{itemize}

	\item{\textbf{Goal 1:}\textit{ Develop Improved Model-Based Representations of Mobile Apps:} Current approaches for modeling the behavior or user interface of mobile apps fail to capture all program attributes necessary for enabling effective GUI-based testing. Thus, we aim to explore modeling techniques that lead to a unified program model which takes into consideration information from the GUI, typical usage scenarios, expected (or unexpected) contextual states, and typical faults that befall mobile apps.}

	\item{\textbf{Goal 2:}\textit{ Expanding Automated Testing Goals:} Current techniques for mobile GUI testing are chiefly concerned with a common goal, called \textit{destructive testing}, which attempts to ``stress test'' and application to uncover faults.  However, little attention, has been given to automating other types of testing goals, such as use case-based testing, which would automatically generate tests around specific use cases to help uncover software regressions. We aim to explore and evaluate~\cite{Linares-Vasquez:FSE'17,Moran:ICSE'18a} automated approaches that facilitate different testing goals.}

	\item{\textbf{Goal 2:}\textit{ Mining Software Repositories and User Reviews to Drive Automated Testing:} The wealth of information in software repositories and user reviews of mobile apps can used to drive automated testing techniques toward more beneficial outcomes. For example, using user reviews to help guide testing toward potential bugs. Thus, we aim to leverage this information to inform next-generation automated techniques.}
	
\end{itemize}

\section{Lessons Learned}

\subsection{Lesson 1: Develop a Strong Sense of Persistence}

	For every successful project, there may be several failed ideas or scientific explorations. The most successful academic researchers are often those that have a relentless sense of persistence and dedication when it comes to their work. They will always be willing to try a new approach, or attack a problem from a different angle, and rarely get discouraged by negative initial results or findings. This sense of drive and dedication is a key facet of becoming a successful researcher.

\subsection{Lesson 2: Follow your Research Passion}

	People tend to do their best work when they are truly passionate about the task at hand. This also tends to be true in research. If a researcher is working on a project they are truly passionate about, or that they feel might truly have an impact on developers, or the research community, they tend to work harder to ensure the success of that project. Therefore, find the topics that truly excite you, or that you feel have the potential for a large impact, and pursue them relentlessly.

\subsection{Lesson 3: Collaborate with Industry}

	In the course of the dissertation work presented in this paper, we were lucky enough to collaborate with industry professionals in developing approaches aimed to help aid software developers and designers. Such collaborations are invaluable, as they help to (i) ground research in reailty, (ii) help guide research toward practical problems, and (iii) facilitate research impact. All prospective PhD students in SE should strive toward some form collaboration with industry to reap these benefits. However, such collaborations can be hard to form and sustain. Thus, the following three pieces of advice may be beneficial for future students: (i) make it a point to meet and discuss your work with industry practitioners to establish relationships, (ii) during collaboration, build prototypes, not products, and evaluate them in a way that shows value to collaborators, and (iii) be open to tackling industrial problems.

\section*{Acknowledgment}

The author would like to thank his advisor, Dr. Denys Poshyvanyk for his invaluable guidance throughout his tenure as a doctoral student. He is also grateful to all of his co-authors and collaborators, particularly, Mario Linares Vasquez, Carlos Bernal Cardenas, Christopher Vendome, and Massimiliano Di Penta. The doctoral dissertation described in this paper was supported by National Science Foundation grants CCF-1218129, CCF-1253837, and CCF-1525902.

\balance
\bibliographystyle{unsrt}
\bibliography{ms}

\begin{thebibliography}{10}

\bibitem{Moran:Dissertation'18}
Kevin Moran.
\newblock Automating software development for mobile computing platforms.
\newblock {\em Doctoral Dissertation}, 2018,
  \url{https://www.kpmoran.com/publications/}.

\bibitem{Brooks:Computer'87}
F.~P.~J. Brooks.
\newblock No silver bullet essence and accidents of software engineering.
\newblock {\em Computer}, 20(4):10--19, April 1987.

\bibitem{Moran:ICPC'18}
Kevin Moran, Carlos Bernal-C{\'a}rdenas, Mario Linares-V{\'a}squez, and Denys
  Poshyvanyk.
\newblock Overcoming lanaguge dichotomies: Toward effective program
  comprehension for mobile app development.
\newblock In {\em 26th IEEE International Conference on Program Comprehnsion},
  May 2018.

\bibitem{Moran:ICSE'18}
Kevin Moran, Boyang Li, Carlos Bernal-C\'{a}rdenas, Dan Jelf, and Denys
  Poshyvanyk.
\newblock Automated reporting of gui design violations for mobile apps.
\newblock In {\em Proceedings of the 40th International Conference on Software
  Engineering}, ICSE '18, pages 165--175, New York, NY, USA, 2018. ACM.

\bibitem{Moran:ArX'18}
Kevin Moran, Carlos Bernal-C{\'a}rdenas, Michael Curcio, Richard Bonett, and
  Denys Poshyvanyk.
\newblock Machine learning-based prototyping of graphical user interfaces for
  mobile apps.
\newblock {\em IEEE Transactions on Software Engineering}, 2018.

\bibitem{Moran:ICST'16}
Kevin Moran, Mario Linares-V{\'a}squez, Carlos Bernal-C{\'a}rdenas, Christopher
  Vendome, and Denys Poshyvanyk.
\newblock Automatically {{Discovering}}, {{Reporting}} and {{Reproducing
  Android Application Crashes}}.
\newblock In {\em 2016 {{IEEE International Conference}} on {{Software
  Testing}}, {{Verification}} and {{Validation}} ({{ICST}})}, ICST'16, pages
  33--44, April 2016.

\bibitem{Moran:ICSE-C'17}
Kevin Moran, Mario Linares-V{\'a}squez, Carlos Bernal-C{\'a}rdenas, Christopher
  Vendome, and Denys Poshyvanyk.
\newblock {{CrashScope}}: {{A Practical Tool}} for {{Automated Testing}} of
  {{Android Applications}}.
\newblock In {\em Proceedings of the 39th {{International Conference}} on
  {{Software Engineering Companion}}}, ICSE-C '17, pages 15--18, Buenos Aires,
  Argentina, 2017. {IEEE Press}.

\bibitem{Moran:FSE'15}
Kevin Moran, Mario Linares-V{\'a}squez, Carlos Bernal-C{\'a}rdenas, and Denys
  Poshyvanyk.
\newblock Auto-completing {{Bug Reports}} for {{Android Applications}}.
\newblock In {\em Proceedings of the 2015 10th {{Joint Meeting}} on
  {{Foundations}} of {{Software Engineering}}}, FSE'15, pages 673--686,
  Bergamo, Italy, 2015. {ACM}.

\bibitem{Moran:ICSE'16}
Kevin Moran, Mario Linares-V{\'a}squez, Carlos Bernal-C{\'a}rdenas, and Denys
  Poshyvanyk.
\newblock {{FUSION}}: {{A Tool}} for {{Facilitating}} and {{Augmenting Android
  Bug Reporting}}.
\newblock In {\em {{ICSE}}'16}, ICSE'16, May 2016.

\bibitem{Linares-Vasquez:MSR'15}
Mario Linares-V{\'a}squez, Martin White, Carlos Bernal-C{\'a}rdenas, Kevin
  Moran, and Denys Poshyvanyk.
\newblock Mining {{Android App Usages}} for {{Generating Actionable GUI}}-based
  {{Execution Scenarios}}.
\newblock In {\em Proceedings of the 12th {{Working Conference}} on {{Mining
  Software Repositories}}}, MSR '15, pages 111--122, Florence, Italy, 2015.
  {IEEE Press}.

\bibitem{Moran:MobileSoft'17}
Kevin Moran, Richard Bonett, Carlos Bernal-C{\'a}rdenas, Brendan Otten, Daniel
  Park, and Denys Poshyvanyk.
\newblock On-{{Device Bug Reporting}} for {{Android Applications}}.
\newblock In {\em {{MobileSOFT}}'17}, MobileSoft'17, May 2017.

\bibitem{Ko:ICSE'05}
Andrew~J. Ko, Htet Aung, and Brad~A. Myers.
\newblock Eliciting design requirements for maintenance-oriented ides: A
  detailed study of corrective and perfective maintenance tasks.
\newblock In {\em Proceedings of the 27th International Conference on Software
  Engineering}, ICSE '05, pages 126--135, New York, NY, USA, 2005. ACM.

\bibitem{Roehm:ICSE'12}
Tobias Roehm, Rebecca Tiarks, Rainer Koschke, and Walid Maalej.
\newblock How do professional developers comprehend software?
\newblock In {\em Proceedings of the 34th International Conference on Software
  Engineering}, ICSE '12, pages 255--265, Piscataway, NJ, USA, 2012. IEEE
  Press.

\bibitem{Linares-Vasquez:ICSME'17}
Mario Linares-V{\'a}squez, Kevin Moran, and Denys Poshyvanyk.
\newblock Continuous, {{Evolutionary}} and {{Large}}-{{Scale}}: {{A New
  Perspective}} for {{Automated Mobile App Testing}}.
\newblock In {\em 2017 {{IEEE International Conference}} on {{Software
  Maintenance}} and {{Evolution}} ({{ICSME}})}, ICSME'17, pages 399--410,
  September 2017.

\bibitem{Choudhary:ASE'15}
Shauvik~Roy Choudhary, Alessandra Gorla, and Alessandro Orso.
\newblock Automated {{Test Input Generation}} for {{Android}}: {{Are We There
  Yet}}? ({{E}}).
\newblock In {\em 2015 30th {{IEEE}}/{{ACM International Conference}} on
  {{Automated Software Engineering}} ({{ASE}})}, ASE'15, pages 429--440,
  November 2015.

\bibitem{Joorabchi:ESEM'12}
Mona~Efrani Joorabchi, Ali Mesbah, and Philippe Kruchten.
\newblock Real {{Challenges}} in {{Mobile App Development}}.
\newblock In {\em Empirical {{Software Engineering}} and {{Measurement}}, 2013
  {{ACM}} / {{IEEE International Symposium}} On}, ESEM'12, pages 15--24,
  October 2013.

\bibitem{Kochhar:ICST'15}
Pavneet~Singh Kochhar, Ferdian Thung, Nachiappan Nagappan, Thomas Zimmermann,
  and David Lo.
\newblock Understanding the {{Test Automation Culture}} of {{App Developers}}.
\newblock In {\em 2015 {{IEEE}} 8th {{International Conference}} on {{Software
  Testing}}, {{Verification}} and {{Validation}} ({{ICST}})}, ICST'15, pages
  1--10, April 2015.

\bibitem{Linares-Vasquez:ICSME'17a}
Mario Linares-V{\'a}squez, Carlos Bernal-Cardenas, Kevin Moran, and Denys
  Poshyvanyk.
\newblock How do {{Developers Test Android Applications}}?
\newblock In {\em 2017 {{IEEE International Conference}} on {{Software
  Maintenance}} and {{Evolution}} ({{ICSME}})}, ICSME'17, pages 613--622,
  September 2017.

\bibitem{Linares-Vasquez:FSE'17}
Mario Linares-V{\'a}squez, Gabriele Bavota, Michele Tufano, Kevin Moran,
  Massimiliano Di~Penta, Christopher Vendome, Carlos Bernal-C{\'a}rdenas, and
  Denys Poshyvanyk.
\newblock Enabling {{Mutation Testing}} for {{Android Apps}}.
\newblock In {\em Proceedings of the 2017 11th {{Joint Meeting}} on
  {{Foundations}} of {{Software Engineering}}}, FSE'17, pages 233--244,
  Paderborn, Germany, 2017. {ACM}.

\bibitem{Moran:ICSE'18a}
Kevin Moran, Michele Tufano, Carlos Bernal-C\'{a}rdenas, Mario
  Linares-V\'{a}squez, Gabriele Bavota, Christopher Vendome, Massimiliano
  Di~Penta, and Denys Poshyvanyk.
\newblock Mdroid+: A mutation testing framework for android.
\newblock In {\em Proceedings of the 40th International Conference on Software
  Engineering: Companion Proceeedings}, ICSE '18, pages 33--36, New York, NY,
  USA, 2018. ACM.

\end{thebibliography}

\end{document}